# Mobile Sound Recognition for the Deaf and Hard of Hearing

**Leonardo A. Fanzeres**
(PPGI, DCC/IM, Universidade Federal do Rio de Janeiro, Rio de Janeiro, RJ
leonardofanzeres@ufrj.br)

**Adriana S. Vivacqua**
(PPGI, DCC/IM, Universidade Federal do Rio de Janeiro, Rio de Janeiro, RJ
avivacqua@dcc.ufrj.br)

**Luiz W. P. Biscainho**
(DEL/Poli & PEE/COPPE, Universidade Federal do Rio de Janeiro, Rio de Janeiro, RJ
wagner@smt.ufrj.br)

**Abstract:** Human perception of surrounding events is strongly dependent on audio cues. Thus, acoustic insulation can seriously impact situational awareness. We present an exploratory study in the domain of assistive computing, eliciting requirements and presenting solutions to problems found in the development of an environmental sound recognition system, which aims to assist deaf and hard of hearing people in the perception of sounds. To take advantage of smartphones computational ubiquity, we propose a system that executes all processing on the device itself, from audio features extraction to recognition and visual presentation of results. Our application also presents the confidence level of the classification to the user. A test of the system conducted with deaf users provided important and inspiring feedback from participants.

**Keywords:** Environmental sound recognition, HCI, Hearing loss, Interface accessibility, Mobile computing, Situational awareness, Ubiquitous systems.

## 1 Introduction

Consciousness of what happens in the surrounding environment is strongly dependent on an individual's capacity to perceive sounds and accurately identify events related to them. Acoustic insulation can seriously compromise the ability of a person to acquire situational awareness, which is important for the execution of daily tasks, social interaction and even personal safety. Hearing individuals may not realize how much they depend on auditory ability to perceive what is happening around them. Environmental sound awareness is necessary in a much higher number of situations than commonly imagined. Examples of this problem are provided by Matthews et al. [1], who propose an environmental sound recognition (ESR) system for deaf users. Among other cases presented in their study, a participant reported that once had forgotten the vacuum cleaner on all night, since the device did not provide any visual cues that it was in operation. Based on previous research [2], the present document describes the design, development and test of a mobile ESR system that aims to expand deaf individuals' situational awareness. We present an exploratory study in the field of assistive computing, describing solutions for problems encountered during the development of an ESR system specifically designed for deaf users. Currently there are still few studies on this topic, despite its importance. To foster future works, we formalize the requirements that guided the development of our system and provide details about the implementation of our solution.

In general terms, existing studies on sound recognition are divided into three categories: speech, music and environmental sound. In the latter case, the most common approach is the use of predefined environmental sound classes, which can then be applied to the indexing/retrieval of audio/video documents and in surveillance systems, for instance. In our study, due to the diverse and ever-changing nature of sounds that the system is supposed to cover, we had to address the open-set problem regarding ESR as well as the representation of uncertainty. During recognition tasks, with the system being executed in an environment where there is no control over the occurrence of sounds, results can be quite inconsistent. One alternative to minimize this problem is to provide the user with information on the confidence level for the classification results. To that end, we calculate a fit between the detected sound and the recognized class through the use of a *group pertinence index* (GPI), described in Section 6.

Considering the intended ubiquity of a sound recognition service for deaf users, we propose a solution employing mobile technology. To guarantee service availability, we maintain that the application should perform all processing on the mobile device itself, which imposes certain hardware limitations, affecting the range of sounds that can be handled. No matter how efficient the algorithm, its execution has a computational cost that tends to rise when new sounds are added to the knowledge base (KB). Faced with resource constraints and the wide variety of environmental sounds, we propose the creation of personalized KBs as a way to meet the demand for sound recognition of all potential users without overloading the system.

Our research also includes the release of the VSom mobile app for recording and recognizing environmental sounds, available online for free. Based on results of a previous study [3] on system performance evaluation, we have worked to obtain the best trade-off between sound classification accuracy and computational cost. We also present the feedback obtained from a test of the system with deaf users.

## 2   Related Work

Our review covered preferentially studies that employ mobile technology. However, due to scarcity of works addressing this topic, we also included desktop-based solutions that face the challenge of offering an ESR solution through accessible technology.

Ho-ching et al. [4] developed a prototype that provides two alternatives to visualize sounds: a spectrogram and a map of the environment that indicates the location of sound emitting sources, which was implemented in simulation mode. In their experiment, the user must, from the view that is provided, indicate whether the occurring sound is from a knock at the door or a phone ringing. The results indicate the inference capability of participants when using each of the visualization forms. In a quiet environment the spectrogram alternative provided better outcomes, but during noisy conditions the map performed significantly better. In a overall view, the map resulted in a higher detection/recognition accuracy, 83% compared with 69% for the spectrogram. Nonetheless, participants have shown great interest in this functionality, managing to analyze the spectrogram and recognize other sounds in addition to those employed in the experiment. One disadvantage of using the spectrogram as a way to expand situational awareness is that it demands a lot of attention from the user, which may not be feasible in many places and situations. Localization of sound sources has proved to be very effective in assisting sound recognition, but requires a prior mapping of the environment.

Matthews et al. [1] developed a prototype of an ESR system for deaf users, which was trained in a specific controlled environment, in this case an office. In addition to recognizing sounds in the environment, the system indicates the direction of sound emitting sources. Their study offers an important contribution with details on impressions and preferences reported by participants during an initial test, from which we consider especially interesting some demands regarding:

- reliability: access to the confidence level of the sound recognition result;
- usability: possibility of interpreting sound recognition quickly (*glanceability*) and have a history of recognized sounds.

Furthermore, participants of the study indicated the sounds that they would like to recognize or to be alerted. The system, however, is desktop-based and also has limitations regarding responsiveness due to delay in presenting recognition results.

Azar et al. [5] developed a prototype ESR system for deaf users that indicates the direction of sound emitting sources. Additionally, the system provides basic ASR functionalities. However, as in the aforementioned studies, the project is entirely desktop-based. The system tests included the participation of ten deaf users, who were excited about the experienced results.

Yoo et al. [6] conducted a research on ESR with emphasis on mechanical sounds. The study is based on the fact that this type of sound has narrower relevant spectral peak than spectral valley areas compared to sounds in general, and they propose a distance measure called normalized peak detection ratio (NPDR) to discriminate sounds. The key idea is to enable sound recognition to be focused on the relevant bands. Besides providing acceptable sound recognition in noisy environments (achieving 60% with -10dB noise level), the spectral band separation method also reduces false positives of non-registered sounds. This is possible because the values of the NPDR can be used directly as a confidence measure to determine whether a detected acoustic event is a registered sound or not. The study also included the development of a wearable app for smartwatches.

Lu et al. [7] present a framework for the development of applications based on audio sensing, where sound classification is done in consecutive stages, hierarchically refining the result. In the experiment conducted by the authors, the first stage categorizes the sample audio into speech, music or environmental sound using decision trees and Markov models. In the second stage, it was only provided one classifier for speech sounds to distinguish the sampled audio between male and female voice. As for environmental sounds category, the system employs unsupervised learning, grouping the samples by similarity and prompting the user to provide a label whenever a new sound is detected. However, inside this category, the only inference regards the relevance of the sound, which is based on its duration and frequency. They also present an evaluation of the classifiers' accuracy using labeled environmental sound samples from previous work [8]. The test included the following classes: "walking", "driving cars", "riding elevators" and "riding a bus", which achieved 93%, 100%, 78% and 25% of accuracy respectively. The study, however, is not directed to deaf users, but to the development of user activity tracking applications. Therefore, more than informing which sounds are occurring around, the proposal is to reveal patterns of activity recognized in the audio signal captured by the device. The study offers also a detailed approach to perform lightweight sound processing. An important feature of the system is the ability to process the audio signal and perform sound classification without using additional hardware resources. Regarding environmental sounds recognition, the study proposes a system scalable to the general public through the implementation of an adaptive unsupervised learning algorithm, but the problem of increasing the base size is not addressed. The study does not present information about the application's capability to perform real-time classification with a larger number of classes, while maintaining the verified accuracy.

Hipke et al. [9] present the prototype of a mobile application called AudioVision, specifically designed for deaf and hard of hearing users. They propose a personalizable ESR system, which allows the classifier to be trained according to the acoustic context of the user. The main drawback of their approach is that audio classification tasks are non-real-time processes executed in a remote server. Nevertheless, the app offers a basic real-time sound visualization.

Rossi et al. [10] present a real-time mobile ESR application. The system uses a predefined base with six 30-second audio records for each class. Despite the low classification accuracy percentage (approximately 58%), the system offers an important contribution with the proposal of two running modes: an autonomous, in which all recognition process is carried out on the device itself; and a server-based mode, in which the extracted audio features are sent to the server that will run the classification process. The authors also present a comparison between the performances of both running modes using two different smartphone models. After the test they found that the server mode has not shown benefits to justify the overhead of communication. However, in our view, the use of a server's processing resources could provide significant benefits regarding runtime and power consumption if more complex algorithms were used for classification.

Ranjbar and Stenström [11] present Monitor, a vibrotactile aid device for expanding situational awareness of people with severe hearing impairment or deafblindness. The device allows users to perceive environmental sounds by detecting their occurrence and adapting the audio signal to the frequency sensitivity range of the skin. By sensing the generated vibrations, acoustic events can be detected and recognized by the user. With an optional microphone configuration, the direction of the sound source can also be sensed. The device is tested by four participants with Usher's syndrome type I, all born deaf. The reported results show that both alternatives, directional and omnidirectional, can greatly improve participants capacity to perceive sound related events. Besides, after using Monitor, all participants reported music as a new and pleasant experience. The computational part is restricted to the algorithm that translates the captured audio as vibrations. Sound detection and recognition are resultant of participants inference, usually combining different senses. Besides vibrations, they use information from touch, smell, taste, draught (air current), temperature differences and remaining vision.

On their study, Bragg et al [12] provide a web based survey among deaf and hard of hearing participants to determine which sounds they feel most need to be alerted in addition to other preferences regarding sound awareness. The study also included the development of a prototype of a mobile sound detector app and an alpha test describing participants reaction to the experimented functionalities. However, they did not implement any ESR functionality on the app. The performed test simulated real-time recognition through notifications sent manually to the users device.

Jain et al. [13] investigate visualizations for spatially locating sound on a head-mounted display (HMD). Their main goal is to augment deaf and hard of hearing sound awareness during group conversations with partners who use spoken language. They defined eight visual sound feedback dimensions and evaluated them as a design probe with 24 deaf and hard of hearing participants, revealing preferences for specific design options. Furthermore, they implemented a real-time proof-of-concept HMD prototype and conducted an in-lab user study obtaining feedback from 4 new participants. The authors own experience also contributed to enrich the generation of the visualizations since two of them, including the first author, have severe-to-profound hearing loss.

Mielke et al. [14] developed a prototype of an environmental sound detector running on a smartwatch. The design of the application was evaluated with deaf and hard of hearing users employing a simulated sound recognition functionality. The study provided interesting feedback from participants about their preferences on the user interface, such as the suggestion of customizable vibration patterns for sound detection notification. Regarding sound semantics, one participant remarked that it is hard for a prelingual deaf individual to grasp the concept of a sound. Therefore, deaf participants have difficulty to "understand what actually is a sound, which frequencies make up a sound, a unique sound".

In a recent study, Liu et al. [15] developed a mobile app prototype which performs ESR based on deep learning models. High sound recognition accuracy was verified employing a dataset with nine sound classes. The proposed system also presents good performance regarding delay time for sound recognition and battery consumption. Despite that sound recognition is performed entirely on the mobile device, the classifier training is cloud-based and there is no other alternative available on the device since the training of a deep neural network implies heavy computational processes. Additionally, the authors proposed a preliminary solution for addressing overlapping sounds employing unsupervised Non-negative Matrix Factorization (NMF) [16], but it is only available when two or more microphones are provided.

## 3    Adopted Solution

According to a World Health Organization (WHO) estimate, around 466 million people worldwide have disabling hearing loss [1] [2]. Despite the remarkable development of deaf education, there is still space for improvement in the field of assistive computing. The research and development of mobile apps that provide situational awareness expansion for the deaf and hard of hearing can be an important contribution to the functional equalization of this significant group of society.

---

[1] http://www.who.int/pbd/deafness/estimates/en/
[2] Disabling hearing loss refers to hearing loss greater than 40dB in the better hearing ear in adults and a hearing loss greater than 30dB in the better hearing ear in children.

There is a mature range of assistive technologies for deaf and hard of hearing individuals, which has been extensively documented by Hersh et al [17], ranging from body-worn, ear-adapted and implanted devices to various categories of warning and alarm systems with vibro-tactile and visual signals, besides communication systems with specific designs. However, these technologies are frequently expensive and/or can spot the impairment of the user [14], which can seriously compromise their adoption. In addition, the WHO estimates that the current production of hearing aids meets less than 10% of global need [3]. In this context, mobile computing represents a much more accessible alternative and, eventually, with the ability to match or surpass the quality of the service provided by conventional assistive technologies.

Matthews et al. [1] reported specific situations in which environmental sound awareness is necessary. While driving, for example, participants expressed interest in being aware of sounds emitted by other cars and also by their own car. One of them said: "When there is something wrong with the car … it tends to go unnoticed until it is very expensive to fix." It was also related an episode when a couple, whose fire alarm was activated after burning food in the kitchen. Since both were deaf, they did not notice the beep until a hearing friend visited them. These reports are an example of the variety of environments with which it is necessary to deal and also of the service's urgency in certain cases. In the same study, participants reported the need to receive some information about classification confidence level. In a previous work [3] we addressed this problem proposing an equation to calculate the uncertainty in classification results. Extended in the present research and incorporated into the proposed system, this solution is presented in Section 6.

In previous studies with desktop-based ESR systems [4] [1], deaf participants expressed a need to use the system at home, at work and in public. The need to recognize sounds in multiple contexts was confirmed in interviews with professionals from the National Institute for the Education of the Deaf (INES, acronym for *Instituto Nacional de Educação de Surdos*), where they expressed the importance of adopting a solution that provides continuous service anywhere. In addition, a solution designed to address such situations should also have good usability. These observations were considered in the system design so that its use does not interfere, or interferes minimally, with the dynamics in which the user may be inserted. Furthermore, in another study with a desktop based ESR system [5], deaf participants of an initial test showed great interest in a mobile version of the system. Further evidence of the interest for mobile sound detection can be found in [12]. In the present paper, we discuss the use of mobile computing for the development of an ESR system that visually presents the outcomes to the user. Environmental sound recognition through an open-set approach is still a relatively unexplored subject [18] [19], especially when processing is based on mobile technology. We address problems related to system knowledge coverage and dynamicity by proposing the personalization of the KB, described in Section 4.

ESR service availability and ubiquity are also major concerns in our approach. Because of their mobility and processing capabilities, smartphones are the first devices to provide real computational ubiquity [20]. Several benefits may be offered by mobile devices, which has led this technology to be considered a universal service [21]. However, due to resource constraints, mobile applications may become dependent on remote processing, such as cloud computing [22]. This can be a problem, since immersive applications designed to provide real-time information are more susceptible to failure when using services provided remotely. Another dependency that should be avoided is related to the necessity of using resources provided by the so called "smart spaces". Cáceres and Friday [23] emphasized the importance of reducing the need for local infrastructure in ubiquitous systems. The unavailability of such infrastructure can disable the service if the resource is not supplied by the device. Given these circumstances, we propose an ESR system that is not dependent on infrastructure and performs all processing on the mobile device itself, from audio features extraction to classification and presentation of sound recognition results to the user.

---

[3] http://www.who.int/mediacentre/factsheets/fs300/en/

Based on the aforementioned considerations, in addition to issues presented in the Introduction, we have formalized the requirements that guided the development of our ESR system, presented in Table 1.

| | Requirement | Description |
|---|---|---|
| R1 | Service availability | The system must perform the whole recording and sound recognition process using only resources of the detector device, without relying on remote processing or information, either by network or any other connected device. |
| R2 | Ubiquity | The system must be usable at home, at work, and in public places such as inside a bus, a restaurant or a classroom. |
| R3 | Adoptability | The system must be compatible with affordable and widely accessible technology. |
| R4 | Knowledge coverage | The system must be able to address the open-set problem, recognizing environmental sounds of any type. So there is no possibility of assuming a prior structure of the sounds to be recognized. In addition, the system must also be capable of processing knowledge bases large enough to meet the knowledge coverage demanded by the user. |
| R5 | Knowledge dynamicity | The system must use a knowledge base that supports the addition and exclusion of sound classes dynamically, thus allowing its update. |
| R6 | Responsiveness | Sound recognition time, from audio signal processing to classification results, should be as short as possible, not exceeding 5 seconds, which was verified in previous studies [24] as being the maximum waiting time tolerated by mobile devices users. |
| R7 | Reliability | Considering the technical difficulties involved in an open-set approach on environmental sounds recognition, it is important that, besides obtaining high accuracy in classification tasks, the system presents the recognition results together with some measure regarding its confidence. |

| R8 | Informativity | The main informative function of the system is sound recognition, but it must also provide real-time visualization/perception of sounds that are being captured or played employing lower-level descriptions of the audio signal. In other words, the system must provide a feedback of the detected sound without the interpretation of classifiers, indicating, for example, the intensity and/or frequency of the audio signal in a form that can be easily interpreted by the user. This information is critical for the deaf user to know precisely the moment when sounds occur and are recorded, recognized or played. |
|---|---|---|
| R9 | Usability | User interaction should not require much attention for the interpretation of recognition results and also leave user's hands free. Furthermore, the system must provide a visual and/or haptic alert function to the user about the occurrence of relevant sounds in the environment. The system should also provide the history of recognized sounds, so that the user has the alternative to check sounds that were recognized in the moments when he/she was not looking to the results display. The main purpose of this requirement is to allow the use of the system without compromising other tasks that may be being carried out by the user. The system should also prioritize the use of images in the graphical interface, thus trying to work around prelingual deaf individuals problems with written language interpretation [25]. |

*Table 1: System requirements. System requirements*

## 4  The Knowledge Base

The construction of a knowledge base (KB) for an ESR system such as the one we propose brings up specific problems regarding knowledge coverage (R4) and dynamicity (R5). In this section we present our solution to address such issues.

We consider an environmental sound to be any detectable acoustic event which produces an audio signal in the frequency range between 20 and 24,000 Hz. Essentially, these are detectable sounds within the limits of human hearing including a wide margin in the higher frequencies, considering air conducted signals. However, depending on the device's built-in microphone, the frequency range of sounds detected by the application may vary. Speech and music can be included, but considering only the recognition of corresponding classes. For example, in the case of speech, sounds can be classified as deep or high-pitched voice and may also indicate whether it is a low voice or a scream. Automatic speech recognition (ASR) and music identification are not addressed in this research, however, the information provided by the proposed solutions can assist future studies which aim to address such functionalities using mobile technology. Moreover, our work does not address problems related to recognition of overlapping sounds.

In recording tasks, the user classifies the sound by setting its name on the class attribute. This is the attribute which will be inferred and provided back to the user during sound recognition processes. In order to compare and classify data efficiently, the audio description consists of feature vectors, obtained through a process called audio feature extraction [26].

| Audio features | # of values |
|---|---|
| Spectral Rolloff Point [27] | 1 |
| Spectral Flux [27] | 1 |
| Standard Deviation of Spectral Flux | 1 |
| Compactness [27] | 1 |
| Spectral Variability [27] | 1 |
| MFCC [26] | 13 |
| LPC [26] | 9 |

*Table 2. Composition of the knowledge base. The final feature vector dimension is 54, which includes both average and standard deviation values from each feature. Nominal attributes "class" and "environment" are also recorded with each sound record.*

The audio description allows automatic sound classification, which will be performed in the recognition process. Besides the sound class and the environment where it was recorded, the sound instances are represented by a feature vector, as showed in Table 2. The features final values are the average and the standard deviation of the values extracted from each original signal window with a duration of approximately 23ms. A Hann function (*hanning*) [26] is applied to smooth windows edges, thereby minimizing the effect known as spectral leakage [28] during fast Fourier transform (FFT). It was not necessary to overlap windows because there is no need to re-synthesize the signal. Configuration of signal capturing is the following: sequence length of 0.4 to 2.7 s; sampling rate of 48,000 Hz; 16 bit depth; mono channel.

The required knowledge coverage (R4) prevents the use of some known sound recognition approaches, as it implies an open-set approach of ESR. For speech or music recognition, classification processes employ predefined structures that can be updated from time to time following the evolution of addressed sounds. Similarly, existing studies on ESR typically employ predefined sound classes and are usually applied to the indexing and retrieval of audio and video documents. In such cases, the subject is approached in specific and/or monitored situations, relying on the support of pre-structured information, which means the system can be designed to address only a specific set of sounds. In an open-set approach of ESR this strategy is not feasible due to the wide variety of sounds included [29], especially if we consider the limitations of processing resources on mobile devices (R1) and the time required for the presentation of sound recognition results (R6). To address this problem, we adopted the strategy of personalizing the KB, allowing users to build the structure of their own sonic universe. This approach seeks to satisfy the system requirements of knowledge coverage (R4) and dynamicity (R5). Using personalized data, only information related to user-defined sounds are processed, reducing the computational cost of classification and avoiding the need to scale the system with additional resources by using cloud computing e.g., which would compromise the service availability requirement (R1). However, in this case, personalization implies entrusting the user with the construction of the KB, which may restrict the universe of sounds known by the application to the user's sampling capacity. As described in Section 8, deaf users can perform sound sampling tasks, but, in most cases, their capacity is limited to sounds related to some visual manifestation or sounds that can be generated by themselves. For other sounds, there is also the possibility that users employ the spectrogram available in the application as a reference of occurring audible events, but this would require a lot of practice to get reliable samples.

## 5  System Performance

In this section, we summarize an experiment [3] we conducted while in the prototyping phase of the system. The study analyzed the performance of classification algorithms using KBs generated on mobile devices. The results helped us define parameters for the development of a system compatible with a dynamic KB, as required in R5. Classifier performance tests were carried out applying 10-fold cross-validation with a KB composed by instances corresponding to 300 audio records equally distributed in 30 classes. In addition, tests with smaller bases were performed to verify the behavior of classification when the size of KBs varies. The experiment employs feature vectors with the composition presented in Table 2. In order to simplify the experiment, classification tests were performed on a Dell Vostro[4] 3500 notebook with Intel Core[5] i3 processor at 2.40GHz × 4 with 4GB of RAM, running on the operating system Ubuntu[6] 13.10 32-bit. To ensure consistency and reliability, tests were conducted using only KBs built with the mobile application prototype and running the same classification algorithms implemented on it. We also conducted tests to verify the duration of classifiers training and sound recognition processes when executed in mobile devices. The employed KBs are available online[7].

Sampling tasks were carried out in different locations, most of them in home environments. Other locations included parks, restaurants, parking lots, public roads or beaches. In most cases, the sampling was performed with relative control over audio capture quality, but never in a recording studio. This means samples were not completely clean, which would lead to unrealistic outcomes during classification tests. Since the purpose of the experiment was not to verify the system usability, but the accuracy of sound classification, samples were recorded by a hearing user. The employed 30 sound classes are the following: Birdsong; People talking; Deep voice; High-pitched voice; Whistle; Snoring; Knocking; Door lock; Cough; Sneeze; Clapping; Keys jingling; Shushing; Rummaging in plastic; Vacuum cleaner;

---

[4] Dell and Vostro are trademarks of Dell Computer Corporation.
[5] Intel and Intel Core are trademarks of Intel Corporation.
[6] Ubuntu is a trademark of Canonical.
[7] http://purl.org/vsom/experiment_1/kbases

Washing machine; Washing machine spinning; Microwave oven; Hair dryer; Phone ringtone; Pressure cooker; Shower; Ocean waves; Faucet; Fountain; Escaping air; Cars passing; Heavy vehicles passing; Garage warning sign; Crossing guard whistling. However, this is not a permanent fixed set, since the proposed application allows classes to be added or deleted according to the user's need.

To compare different classifiers, we chose four algorithms: nearest neighbor based on Euclidean distance, two employing probabilistic graphical models and one using decision trees. For the nearest neighbor algorithm, the resultant class will be the class of the instance closest to the tested instance. As to probabilistic methods, we used naive Bayes and Bayes network classifiers, where instance attributes, originally numerical, are converted to nominal attributes by supervised discretization [30]. Finally, an *ensemble* of random decision forests was also tested: decision trees have excellent performance as to classification time, but the accuracy is often lower. The ensemble tries to compensate for the lower accuracy of this method by applying a sequence of iterations with decision forests. The obtained accuracies are the following: Nearest neighbor 92.7%; Naive Bayes 88.7%; Bayes network 89.7%; Random forests 92.3%.

During the study we also investigated classifiers behavior with different KB sizes (Fig. 1). The results were used as a benchmark for setting the minimum number of samples per class required by the application to keep acceptable accuracy. These tests also provided information on potential KB coverage by investigating accuracy behavior when the number of classes increases. Specifically, over 24 classes, the decreasing percentage of correct classification tends to approach a horizontal asymptote, i.e. to stabilize at a fixed value in the vertical axis. This result indicates that the addition of sound classes should not imply significant loss of accuracy. However, tests with larger KBs are still needed to confirm this tendency.

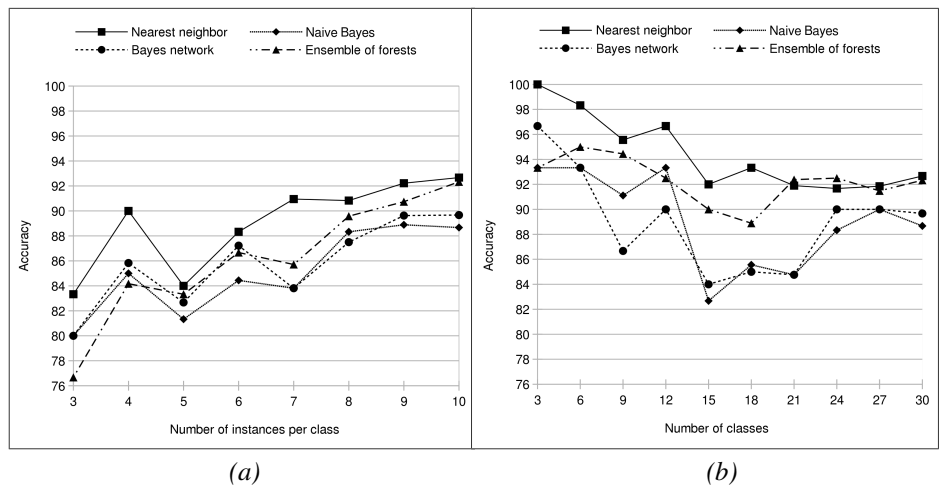

*Figure 1: Classification behavior. (a) Increasing the number of instances per class. In all points on the graph the base has 30 classes. (b) Increasing the number of classes. In all points on the graph each class has 10 instances.*

To evaluate the application responsiveness (R6), we carried out tests on a mobile device to verify the duration of classifiers training and sound recognition (2). Both processes employed the complete KB with 300 instances. Sound recognition includes audio features extraction and classification, which are tasks initiated asynchronously and synchronized later. Therefore, the measurement of running time was conducted evaluating the process as a whole, i.e., the time is measured from the beginning of features extraction to the end of classification. The device used has model reference Xperia[8] C1604 and 1 GHz processor with Android 4.1.1 operating system. As for the classifier training time, it only interferes in the application usage when the KB is modified, because in this case the classifier needs to be re-trained. The training process is also executed when recognition functionalities are activated for the first time after the application launch.

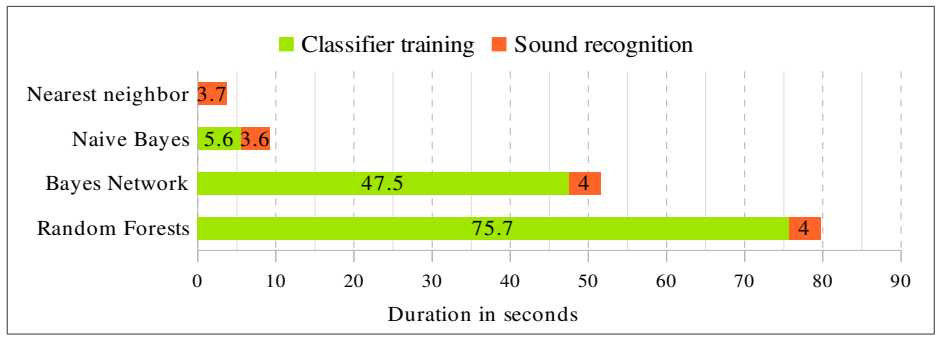

*Figure 2: Classifier training and sound recognition duration*

Despite having obtained the lowest accuracy around 89%, classification with naive Bayes proved to be a good alternative for the application. One advantage over nearest neighbor algorithm, which presented the highest accuracy, is that part of the processing can be advanced during the training phase. Additionally, training with naive Bayes was much faster than in options with neural networks and decision trees. This great difference in computational cost combined with satisfactory accuracy indicates the possibility of using larger bases, what is essential to meet the required knowledge coverage (R4).

## 6    Classification Uncertainty

Considering the tests that verified classifiers performance presented in Section 5, the instances have features generated from a known universe of sounds, i.e. a closed-set approach. Learning is supervised and recordings are carried out and labeled by the user. These tests evaluate the consistency of classes, the selection of features and the classifier accuracy, but not the application's ability to effectively recognize sounds in real situations, i.e. an open-set approach. During the recognition process, sounds not known by the application, i.e. sounds belonging to classes not present in the KB, are also classified and results in this case can be quite inconsistent. In this section we extend a previous proposal [3] regarding classification uncertainty. This problem was

---

[8]  Xperia is a trademark of Sony Mobile Communications AB.

also addressed during the development of another ESR system for deaf and hard of hearing [31], in which authors proposed that three confidence levels be presented to the user. These levels were based on thresholds established using test data obtained from a sound classification system developed by Robert Malkin [32]. During the interview with deaf users conducted by Matthews et al. in the aforementioned study, participants expressed their concerns about the ESR system showing false positives. One participant said "I am worried about it showing 'voices' when I'm at home. If no one was there I would wonder 'what is going on?'" To address this issue, we chose to provide a feedback to the user presenting a classification confidence level based on an equation we developed, which aims to calculate what we call *group pertinence index* (GPI). Essentially, the GPI returns a value that indicates how close a detected sound is to the recognized sound class. Given a new instance, the GPI indicates its degree of pertinence to the group formed by the instances of the resultant class. The calculation of the index, shown in Equations (1), (2), (3) and (4), does not represent a significant increase in the total computational cost of sound recognition, since only the instances pertaining to the resulting class are used. The equations are formulated from a $P_{m \times n}$ matrix where each line corresponds to one of the $m$ instances of the recognized class, and each column corresponds to one of the $n$ dimensions of the respective attribute vectors (which can be seen as points in $\mathbb{R}^n$). We define:

$d(x,y)$ as the Euclidean distance between the points $x$ and $y$
$c$ as the centroid of $P$
$a$ as the point corresponding to the feature vector generated from the detected sound
$p^*$ as the closest point to $a$, considering Euclidean distance
$g(a,P)$ as the pertinence index of $a$ to the group of points in $P$

$$d(x,y) = \sqrt{\sum_{i=1}^{n} (x_i - y_i)^2} \tag{1}$$

$$c_j = \frac{\sum_{i=1}^{m} p_{ij}}{m}, j=1,2,3,\ldots,n \tag{2}$$

$$d(a, p^*) \leq min(d(a, p_i)), i=1,2,3,\ldots,m \tag{3}$$

$$g(a,P) = d(a,c) - d(p^*,c) \tag{4}$$

## 7  Mobile Application

In this section we present the VSom application version 1.0.2, which was recently released to the public. This version is newer than the one used in the user test presented in Section 8. Apart from the application functionalities and characteristics described here, this version provides a tutorial with an overview of the system operation.

### 7.1  Main Functionalities

The application consists of four main functionalities. Sound recording and manual recognition functionalities provide a spectrogram (Fig. 3a) automatically activated during audio capture, thus allowing the visualization of the signal being processed.

The green part of the spectrogram corresponds to the signal being only captured, while the red part corresponds to the segment being recorded or recognized. The aim of offering this visual feedback is to allow the user to interact during recording and recognition tasks more accurately. The spectrogram can also be activated by simply touching on its image, which allows the user to see the spectrum of sounds occurring in the environment even when no recording or recognition processes are running.

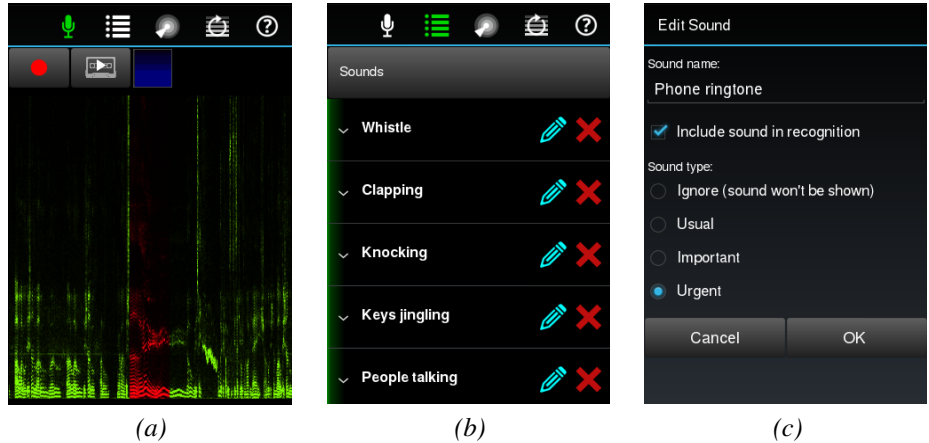

*(a)*          *(b)*          *(c)*

*Figure 3: (a) sound recording. Sound exploration: (b) list of sounds. (c) sound editing form.*

*Sound Recording* – This functionality, presented when the application launches (Fig. 3a), enables users to build their own sound collection. The process is triggered by the user, but the recording is effectively started only when a sound is detected. Signal capture stops automatically when the end of the sound is detected, but the user has the option to terminate the process at any time. Further details are presented in Section 7.6. Fig. 4 shows the process corresponding to this functionality. When the recording ends, the user is prompted to label the sound, i.e. set the corresponding class, and optionally indicate the environment in which the recording was made.

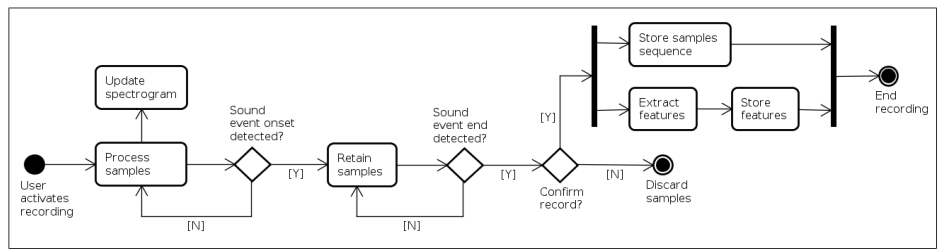

*Figure 4: Model of the sound recording process in a UML activity diagram*

*Sound Exploration* – This functionality allows users to browse their data. The corresponding screen is shown in Fig. 3b, where a list of stored sounds is presented. By clicking on a sound item, a drop-down sublist of its respective records is displayed. Then the user can choose a record and play back the audio, whose frequencies are displayed in a spectrogram. The sound list also allows the user to edit or delete its items. The sound data editing form (Fig. 3c) allows, among other things, to specify a meta-category that differentiates the sound in four levels of importance: irrelevant (ignore sound), usual, important or urgent. This will apply a filter to the results and can trigger custom alerts in the automatic recognition functionality. It is also possible to list sound records grouped by the environment in which they were captured. Similarly, environments can also be edited or deleted.

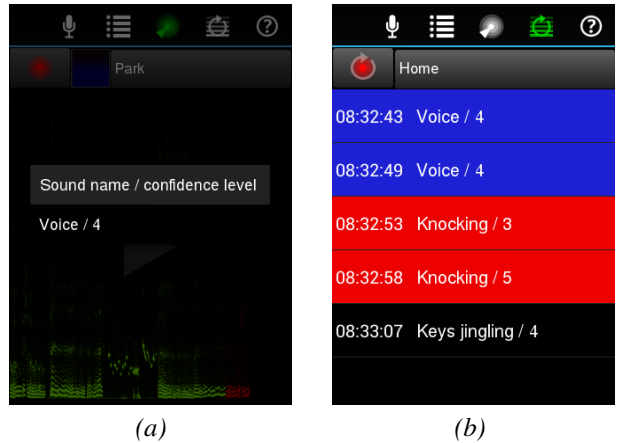

*(a)                                (b)*

*Figure 5: (a) manual sound recognition. (b) automatic sound recognition*

*Manual Sound Recognition* – This functionality allows the user to set the moment the recognition task is performed. After the process is triggered, a spectrogram is displayed for assisting the user to identify the moment at which the sound occurs. Optionally, the user can set the environment to improve the accuracy of sound classification. Once recognition is complete, the result is presented and the process ends. As shown in Fig. 5a, in addition to the recognized sound label, the result also reports the classification confidence level. Fig. 6 shows the process corresponding to this functionality.

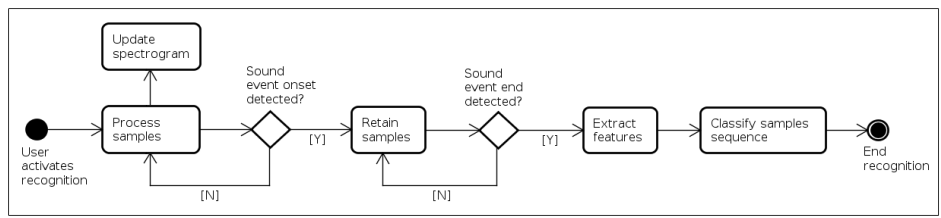

*Figure 6: Model of the manual sound recognition process in a UML activity diagram*

*Automatic Sound Recognition* – This functionality provides a real-time updated list of occurring sounds. Each list item displays the time the sound was detected, the recognized sound label and the confidence level of the corresponding classification. The list also notifies the user about sound importance, which can be previously set in the sound data editing form (Fig. 3c). As shown in Fig. 5b, the application presents black background for usual sounds, blue for important sounds and red for urgent sounds. When importance is set to "ignore", the respective sound remains included in the recognition process, but it is not displayed when it is recognized. Fig. 7 shows the process corresponding to this functionality, which runs continuously in separate threads for each sound detected in the task "Process samples" until being stopped by the user.

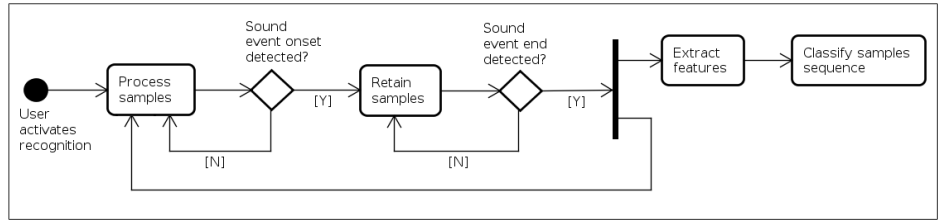

*Figure 7: Model of the automatic sound recognition process in a UML activity diagram*

### 7.2 Implementation

Due to required adoptability (R3), the application was developed to run on the Android operating system from version 2.3.3, enabling its installation in all devices currently operating with Android worldwide[9]. The implementation uses the Java programming language. As specified in the service availability requirement (R1), the entire processing cycle is executed on the mobile device itself, which includes execution from signal capture to spectrogram display, audio features extraction and sound recognition. In addition, two open source libraries contributed significantly for the efficiency of the application:

- jAudio [27] – implemented in Java, the jAudio library provides the classes responsible for audio features extraction. We adapted the library to the Dalvik virtual machine, deployed on devices with Android operating system.
- Weka [33] – data mining library with several statistical methods implemented. Since version 3.0, released in 1999, the Weka is available entirely in Java, executable on desktop devices with Java virtual machine installed. In the application featured here it was used a version developed by RJ Marsan[10], which has also been adapted to the Dalvik virtual machine.

### 7.3 Classification Algorithm

After reviewing results of the classification experiment described in Section 5, the application was configured to use a naive Bayes classifier, which provided the best trade-off between classification accuracy and computational cost. No reinforcement function dependent on user evaluation was used in the classifier training. The user only interferes in classification by adding or removing sound records from the base. Furthermore, the application provides the option to temporarily exclude a sound class from the recognition process, as shown in Fig. 3c.

---

[9] https://developer.android.com/about/dashboards/
[10] https://github.com/rjmarsan/Weka-for-Android

### 7.4 GPI and Confidence Measure

Through the approach presented in Section 6, GPIs with negative values indicate a high degree of pertinence to the group. As to positive values, for more refined results, intervals should be specifically adjusted to each class. Despite that, by using the KB employed in the experiment (Section 5), we found that values between 0 and 1 indicated an average degree of pertinence, while values greater than one occurred mostly in wrong classifications. Based on these observations we have established the thresholds shown in Table 3. The confidence level is then reported to the user along with the resulting class, providing support to the sound recognition process. In the current implementation, the zero level has been set to indicate an unrecognized sound. For levels from 1 to 5, the number is displayed along with the recognized class label, with 1 being the lowest level of confidence and 5 the highest.

| Level | Thresholds |
|---|---|
| 5 | $g(a,P) < 0$ |
| 4 | $0 \leq g(a,P) < 0.5$ |
| 3 | $0.5 \leq g(a,P) < 1$ |
| 2 | $1 \leq g(a,P) < 1.5$ |
| 1 | $1.5 \leq g(a,P) < 2$ |
| 0 | $2 \leq g(a,P)$ |

*Table 3. Classification confidence levels*

### 7.5 Responsiveness

A primary concern on interface design was to ensure satisfactory responsiveness. The processing required to perform features extraction and sound classification may easily block the interface for a couple of seconds. The implementation of asynchronous tasks solved most of the problems, but not all processing can occur in the background, since some functions must provide information to the user in real-time. One such example is the spectrogram, that displays spectral information about audio segments at intervals equivalent to fractions of a second. Each column of the image is generated from half of the 2048 values from the FFT computed over the same number of samples of the original signal captured at 48,000 Hz. For performing FFT, it is necessary to use vectors with dimension equal to a power of 2. We used only half of the values because the spectral information is redundant, therefore the vectors size is 1024. We considered this dimension the best trade-off between time and frequency resolution. Besides, the choice of this dimension for displaying the spectrogram also has taken into account the adequacy of the image to the mobile screen size, considering devices with small and/or low-resolution screens. The quality of the spectrogram image is user configurable as high, medium or low, showing respectively 23, 12 or 8 columns per second, approximately. The purpose of this configuration is to allow the user to adjust this function to the device's processing capacity, thus avoiding noticeable delays between the actual occurrence of sounds and their spectral display. Taking into account the deaf user condition, who cannot aurally check such delays, the application warns their occurrences in recording and manual recognition functionalities turning the blue square into orange/ocher (Fig. 8) depending on the delay.

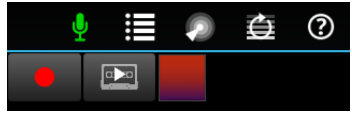

*Fig. 8. Spectrogram delay warning*

Another important issue concerning responsiveness is the time required to perform sound recognition. With naive Bayes classifier, the recognition process, which includes features extraction and classification, lasted for 3.6 seconds. This result is within the 5 second limit established in R6, which refers to the maximum waiting time tolerated by mobile device users [34]. However, this limit is only a benchmark for the development of a system with satisfactory responsiveness. As mentioned in R6, the time spent on sound recognition should be the shortest possible. An excellent performance would be achieved if the recognition result were presented immediately after an acoustic event is detected.

### 7.6 Sound Detection

Automatic sound detection is the system's capacity to detect the start and end of occurring sounds in the environment. During recording and manual recognition of sounds, this feature is especially necessary due to the following issues:
- eventual delay in responding to taps on the device screen, what could impair the manual marking of the start and end of sounds;
- difficulty of manually capturing brief sounds like snaps or bumps;
- difficulty for deaf users to perceive the start and end of sounds that are not associated with any visual or haptic manifestation.

Furthermore, in automatic recognition, detection prevents the application from classifying audio sequences without relevant sounds, such as silent segments, low intensity white noise or distant subtle sounds. In the present application, we have implemented the detection method used by Lu et al. [7] in the development of their framework for systems with audio sensing functions. The method, called by the authors *frame admission control*, is based on the amplitude and spectral entropy of the signal: a low amplitude signal indicates silent segments or low intensity sounds; a high spectral entropy indicates a flat spectrum (e.g. silence or white noise) while a low spectral entropy indicates a well defined pattern on the spectrum. Therefore, spectral entropy and amplitude are attributes that complement each other to determine whether a frame, i.e. a signal segment, will be admitted or not. The admission of the frame will occur if amplitude or entropy reaches the defined thresholds. As for amplitude, the threshold sets the minimum value, whereas for entropy the maximum value is set. With this method, high intensity sounds will be accepted regardless of the entropy because the amplitude is high. Thus a high intensity white noise (e.g. a strong wheezing) will be admitted as a sound. On the other hand, low intensity sounds will only be accepted if they have low entropy, which indicates that are sounds with a well defined pattern on the spectrum as, for example, a person speaking low or distant from the device. However, a low intensity white noise (e.g. air vent sound) will not be admitted because it presents low amplitude and high entropy. The signal amplitude is obtained from the root mean square (RMS) of the $n$-samples sequence, whose calculation is shown in Equation (5). Regarding the entropy, after the FFT is computed from frame $q$, we obtain the spectral magnitude expressed in a vector of dimension $n$, which is treated as a probability density function $p$. Finally, the spectral entropy $H$ is calculated as shown in Equation (6).

$$x_{rms} = \sqrt{\frac{1}{n}\sum_{i=1}^{n} x_i^2} \qquad (5)$$

$$H_q = -\sum_{i=1}^{n} p_i \log p_i \qquad (6)$$

## 8  User Feedback

In order to obtain feedback from potential users, a test with deaf individuals was conducted using an alpha version of the application. The meeting was organized by members of the Projeto Surdos-UFRJ: Prof. Vivian Rumjanek, Prof. Flavio Eduardo P. da Silva and professor and interpreter Tiago Batista, and took place at the Sciences Didactic Laboratory for the Deaf (LaDiCS - Laboratório Didático de Ciências para Surdos). The group dynamics was entirely assisted by Portuguese-LIBRAS (acronym for Brazilian Sign Language in Portuguese) interpretation. It lasted about one hour and consisted of three parts: presentation, testing and application evaluation. Four participants attended the test. Their degree of hearing loss ranged from moderate to profound (as declared by them [11]), and their age from 21 to 28 years. All were familiar with mobile devices. The group was composed of two men and two women. Regarding educational level, one had completed high school; two were students of Bilingual Pedagogy and one already held a degree in that field. All were members of project teams at Federal University of Rio de Janeiro (UFRJ), of which one worked as a children's teacher. The teams and coordinators of these projects are members of the Institute of Medical Biochemistry (IBqM, acronym for *Instituto de Bioquímica Médica*) and have no relation with our department. In addition, participation in the test was voluntary.

During the first part of the meeting, participants were introduced to the system, its purpose and operation. After being familiarized to the proposal of sound recognition, they expressed the necessity of recognizing calls to their names, as well as audible signals from other deaf individuals. They also asked about the possibility of indicating when a particular person is speaking by identifying her/his voice. In addition, they wanted to know if the application could indicate whether the detected voice was from a person who was calm or nervous, happy or sad. Moreover, as they got used to the spectrogram, there was greater interest in the possibility of recognizing sounds by interpreting the generated graphic. Participants also raised questions regarding the amount of sounds that the application could store, since they felt it was important to recognize a wide variety of sounds. Concerned about the application usability in different contexts, they also expressed the need to recognize sounds when the user is on the move. Another issue raised by nearly all participants was the preference for images rather than text to represent sounds in the application. In the second part of the meeting, once they were familiarized with the interface, they performed recordings, played with the spectrogram and tested recognition with sounds they could reproduce like knocking at tables, basic whistling and employing their voice, among other sounds. By the end of the meeting, participants formalized their considerations and evaluated the application by answering the questions posed.

---

[11] Hearing loss degrees employed by the health and education sectors in Brazil [35]: mild: up to 40 dB; moderate: 41 to 70 dB; severe: 71 to 90 dB; profound: above 90 dB.

Besides evaluating the application, participants formalized their priorities regarding situational awareness when answering the first question, in which they were asked about the environments where they felt most need to be aware of the occurrence of sounds. All participants stated that their homes were an important location, where they needed to be alerted. In such environment, especially when they are alone or among other deaf individuals, they will have a greater need to be aware of relevant sounds. Public spaces such as airports, bus stations, street, and also transportation means such as buses, cars or the subway came second. In such environments, they may find themselves alone or among unknown people and therefore be in need to be alerted about stops or imminent dangers they might not perceive. Offices were also considered a priority by most participants, where situational awareness is important to follow ongoing activities and for social interaction. Other locations came next, such as restaurants, pedestrian streets, supermarkets, or classrooms. These locations generally involve less stressful situations, as there is less need to be alerted to a bus/subway stop or some imminent danger. We should also point out that, in classroom environments, they are likely to be among other deaf students, with the presence of a LIBRAS fluent teacher or an interpreter. Participants also commented they would like to be able to recognize specific acoustic events such as: sounds of animals, rain, thunder, wind, music, clapping, gunshots, cry, horns and screams.

In another question, we asked about their perception of the effective situational awareness that the application provided. Despite some negative feedback, three of the participants stated that the application helped them perceive, at least once, what was happening around. Participants were also asked to evaluate the four basic application functionalities, the spectrogram and playback of records. A visual guide with the screens and names of functionalities was provided to participants as a memory aid. Despite a negative evaluation regarding recording and playback functions, the overall appreciation expressed a clear approval. All of the functionalities received positive evaluations (considered Good/Very Good) by most participants, except the manual recognition, which tied between Good/Very Good and Regular.

Furthermore, participants pointed out some of the difficulties they had, particularly with interpreting the spectrogram, and they also found the interface a bit confusing. In the last question, we suggested potential improvements of the application which were evaluated by them. All participants considered important that the application emits a vibratory signal when recognizing a sound, and that they are able to save spectrogram images. Besides, some improvements regarding the app graphic interface were appreciated, such as: present more flashy visual signals when recognizing a sound; provide a higher quality spectrogram and enable the user to customize the app colors.

Moreover, they would also like to be able to save the history of recognized sounds and to know the direction of the sound emitting source. Regarding the application responsiveness, one participant considered unnecessary to recognize sounds faster, while the others opined between Useful and Important on this improvement.

Given participants' feedback, we can reaffirm the relevance of expanding situational awareness for deaf individuals. However, in order to provide solutions for all demands reported in the meeting, further work on the system audio processing capabilities will be necessary, especially regarding ASR and speaker identification. User feedback revealed some weaknesses of our proposal, but, from a general overview, participants demonstrated a positive perception of the application.

## 9   Conclusion and Future Work

ESR systems using mobile computing are a relatively unexplored subject, especially when addressing specific audiences, as is the case of deaf and hard of hearing. We believe that the formalization of requirements and the presentation of solutions to the difficulties found in the development of a system of this nature constitutes an important contribution to the field of assistive computing.

Based on the information provided by our previous study [3], it was possible to guide the application development in order to meet the required responsiveness and service availability with a solution based on mobile computing. The system's requirement for performing the entire recognition cycle on the mobile device influenced many decisions throughout the research. Among them, the alternative for the KB personalization, which, by reducing the classification computational cost, minimized difficulties related to knowledge coverage. With this approach, only sounds included by the user need to be processed, thus classification will be performed over a smaller set of classes. However, due to users' profile and to the extremely diverse and constantly changing nature of environmental sounds, the construction of the KB is difficult to be carried out. Although not addressed in this research, KB sharing between users could solve limitations related to the option for personalization. Hipke et al. [9] had already suggested in their study the use of crowdsourcing to foster the construction of their application KB. After all, according to them, some sounds can be especially difficult to be captured for most users like breaking glass, siren, horn, baby cry, etc. The authors also expressed concern about the use of shared sounds in different environments raising the question: can the sound captured by a user meet the needs of other users in different acoustic environments? Apart from this problem, we also consider the possibility of incompatibility when trying to match the generated KBs with the receiving devices, which may have different microphones. Therefore, additional studies are needed to clarify whether the large scale sharing of KBs is a feasible approach.

As part of this research, the ESR mobile app called VSom was created, and made available online for free [12]. The application can be used for future tests on situational awareness expansion with deaf users participation, hearing impaired people and also users of ear protectors, helmets or other equipment that may cause acoustic insulation of the individual. The app is also useful to assist in understanding the research conducted, as processes, interactions and limitations treated in the study are better illustrated by the application, which can be used by the reader to experience its functionalities.

---

[12] VSom is available for download at https://play.google.com/store/apps/details?id=app.vsom

We also documented here the results of an initial test performed with a group of deaf users. Tests of this nature are scarce, especially using mobile computing. According to most participants, the application was able to improve their situational awareness. They also expressed the need to recognize some specific speech sounds, such as calls to their names, which raises new questions on the topic, since the present study does not address ASR functionalities. During the meeting, the application was presented to participants with the help of a Portuguese-LIBRAS interpreter. In order to facilitate future tests, it is important to provide further supporting material such as videos with explanation in sign language about the main functionalities, thus replacing face-to-face presentation. Additionally, in order to provide a more consistent study of the system effective utility, tests gathering a greater number of users will be necessary. It is also important that such studies encourage the application usage over a longer period, thus allowing participants to experience it in different environments, including their homes.

We developed a system that provided the required processing capabilities to perform ESR tasks, yet, the ubiquity of the service also depends on the availability of the information for the user at the time when the sound occurs. In the current application version, its use is indicated only in environments such as home or office due to the need to maintain the device protected from rubs and bumps that may cause interference to the audio signal being captured. In these environments, users may easily leave the application running on a smartphone that will alert them with visual signals. As for the use in public places or in situations that require greater mobility, alternatives using wearable computing would be needed. Smart watches and other smart devices embedded in bracelets and brooches can complement the service provided by alerting the user with vibrational signals and also displaying the recognized sound label in short messages in case of devices that offer visual output. However, the use of such devices was not addressed in this study. This limitation prevented the required ubiquity (R2) to be completely satisfied.

Contributions from previous studies allowed us to understand deaf and hard of hearing needs in the use of ESR systems and applications with similar objectives. Based on this information, we were able to specify our system requirements and then apply and extend existing sound recognition techniques, adjusting them to the adopted solution. We believe that our approach tackles the issue in the right direction by taking into account specific necessities related to the deaf user experience, gathering the solution to an effective and publicly available application, which we consider to be the major differential from other approaches. Future studies include the search for solutions to the user overload in KB building and the development of new recognition techniques that consider the complexity of environmental sounds.

**Acknowledgements**

This work was supported by CAPES, CNPq and FAPERJ. The authors wish to thank members of the Projeto Surdos-UFRJ: Prof. Vivian Rumjanek, Prof. Flavio Eduardo P. da Silva and professor and interpreter Tiago Batista, and the four participants of the user test, which greatly contributed to this research.